\documentstyle[prd,preprint,epsf,aps,floats,amssymb]{revtex}

\newcommand{\jjmass}{\mbox{$M$}}
\newcommand{\Et}{\mbox{$E_{T}$}}

\newcommand{\chisq}{\mbox{$\chi^{2}$}}

\newcommand{\ipb} {\mbox{pb$^{-1}$}}
\newcommand{\qstar}{\mbox{$q^{\ast}$}}
\newcommand{\Etmax}{\mbox{$E_{T}^{\rm max}$}}
\newcommand{\Etu}[1]{\mbox{$E_{T}^{{\rm jet}#1}$}}
\newcommand{\modetajet}{\mbox{$\mid \! \eta_{\rm jet}  \! \mid$}}
\def\gev{GeV/c$^2$}   
\newcommand{\pbarp}{\mbox{$\bar{p}p$}}

\input psfig.tex

\begin{document}
\tightenlines
\title{Inherent Biases in the Ansatz Function Method used to Search for 
       New Particles Decaying to Two-Jets in $\bbox{\pbarp}$
       Collisions.}

\author{Iain A. Bertram\\
{\it Northwestern University, Evanston, Illinois 60208}}

\date{\today}
\maketitle
\begin{abstract}
 Searches for new resonant phenomena on top of a continuum spectra
 need to make assumptions regarding the shape of the continuum
 spectra.  The Ansatz function method used in previous searches for
 new particles decaying to dijets is investigated and found to have
 inherent biases that cause the 95$\%$ confidence limits on the signal
 cross-section to be miss-measured by up to 30 to 50$\%$.
\end{abstract}  

 Previous searches for hypothetical particles produced in
 proton--anti-proton collisions which decay to
 two-jets~\cite{UA2wjj,ua2xjj,cdfxjj,cdfxjj2} have utilized a
 procedure where the QCD inclusive two-jet mass spectrum is fitted as
 a linear combination of an ansatz function and a signal line shape to
 determine limits on the production cross section for various
 theoretical particles. The reliability of this procedure depends on
 the ability of the ansatz function to correctly model the mass
 spectrum of the continuum background. This paper describes a test of
 this procedure.

 For this test the QCD inclusive two-jet mass spectrum resulting from
 \pbarp\ collisions at $\sqrt{s} = 1.8$ TeV was simulated using the
 Next--to--leading order (NLO) event generator {\sc
 jetrad}~\cite{jetrad} and the CTEQ4M~\cite{cteq} parton distribution
 function (pdf), a renormalization scale ($\mu$) of $0.5\Etmax$ where
 \Etmax\ is the maximum jet transverse energy ($\Et$) of the event and
 a parton clustering algorithm. For the clustering algorithm, all
 partons within $1.3 {\cal R}$ (${\cal R} = \sqrt{\eta^{2} +
 \phi^{2}}$) of one another were combined if they were also within
 ${\cal R} = 0.7$ of their $\Et$ weighted $\eta$,$\phi$
 centroid~\cite{rsep} where $\eta = -{\rm ln}[{\rm tan}(\theta/2)]$,
 $\theta$ is the polar angle relative to the proton beam and $\phi$ is
 the azimuthal angle. The $\Et$ of the jets are then smeared using
 typical collider jet resolutions \cite{jet_resolutions}. The two-jet
 mass for each event was calculated using the two highest \Et\ jets
 assuming that the jets are massless, using the relationship; $M^{2} =
 2 \Etu{1} \Etu{2} [ \cosh( \Delta \eta ) - \cos ( \Delta \phi )
 ]$. The two leading $\Et$ were required to satisfy the requirements
 $\modetajet < 1.0$, $\Delta \eta = \mid \!  \eta_{1} - \eta_{2} \!
 \mid < 1.6$, and $\jjmass > 200$ \gev . The effects of statistical
 fluctuations in the prediction were removed by fitting to an ansatz
 function with the form:
\begin{equation} A \cdot M^{- \alpha}
 \exp{\left[ 1 - \left(\frac{M}{1800}\right) -\gamma
 \left(\frac{M}{1800}\right)^{2} - \delta
 \left(\frac{M}{1800}\right)^{3} \right]^{\beta} 
 {\mathcal{P}}_{n}\left(\jjmass\right) } 
\end{equation} 
 where ${\mathcal{P}}_{n}\left(\jjmass \right)$ is a polynomial of
 degree $n$. The D\O\ Collaboration has demonstrated~\cite{dijet_mass}
 that this simulation is a good representation of its measured Dijet
 Mass Spectrum.

\begin{figure}
\vbox{\centerline{\psfig{figure=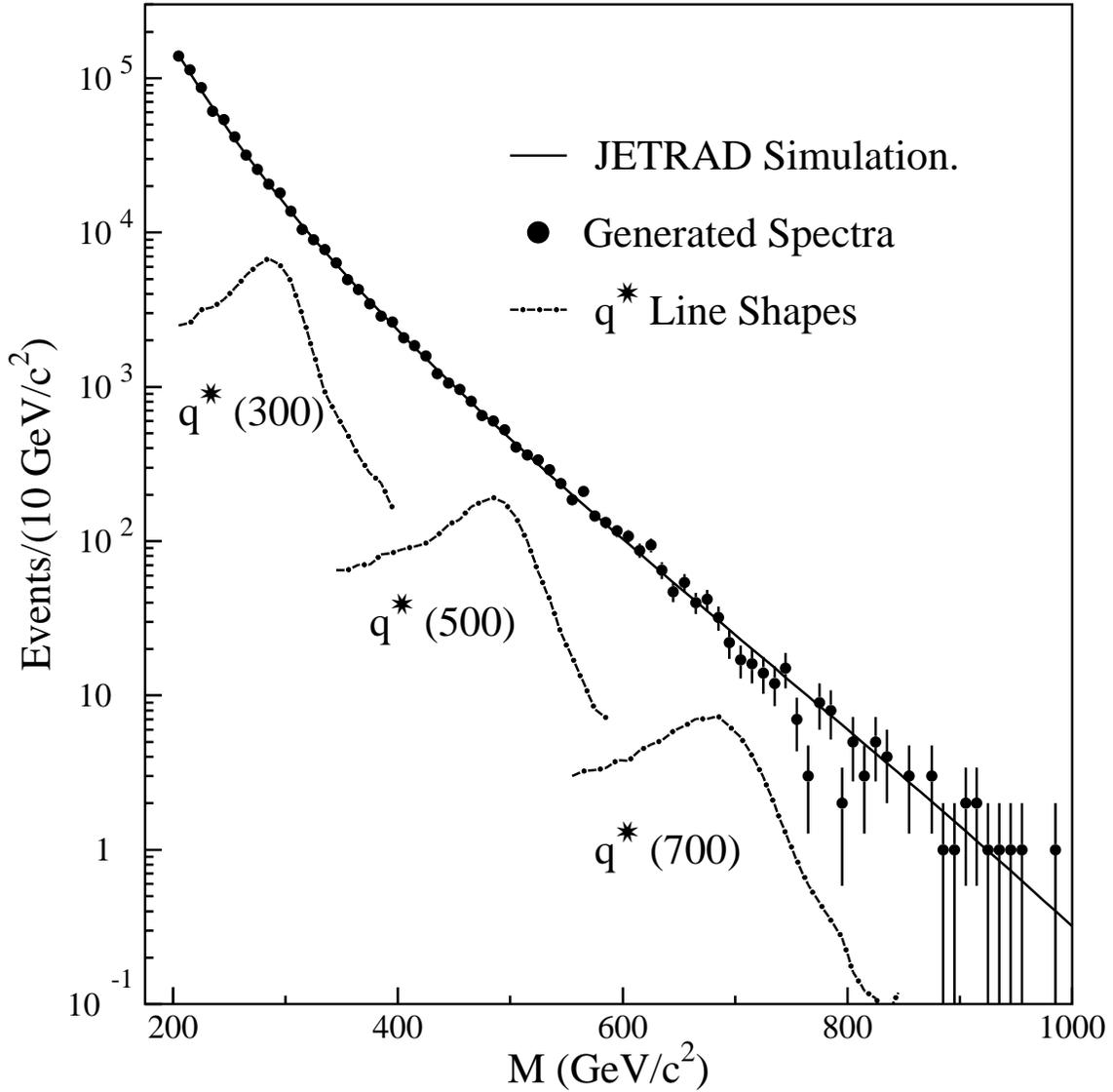,width=6in}}}
\caption{The QCD Dijet Mass Spectrum produced by the {\sc jetrad} program 
(solid line) and one of the randomly generated spectra (solid
circles). Note: That the randomly generated spectra has been corrected
so that all points have the same apparent luminosity. Also shown are
the simulated $\qstar$ at masses of 300, 500 and 700 \gev\ (dash--dot
lines).}
\label{fig_1}
\end{figure}

 To correctly model the statistical fluctuations typical of collider
 data samples, one hundred mass spectra were generated in 10 \gev\
 mass bins by smearing the spectrum obtained above with the Poisson
 fluctuations expected for the cross-section given by the simulation
 per bin times the luminosity for that bin.  Four different mass
 regions, each with a different luminosity, were used to mimic data
 taking conditions, 0.5~\ipb\ for $200 < \jjmass \le 270$ \gev ;
 5.~\ipb\ for $270 < \jjmass \le 370$ \gev ; 50.~\ipb\ for $370 <
 \jjmass \le 500$ \gev ; and 100.~\ipb\ for $\jjmass > 500$ \gev\ .
 An example of one of the randomly generated spectra and the source
 distribution are shown in Fig.~\ref{fig_1}.
 
\begin{figure}
\vbox{\centerline{\psfig{figure=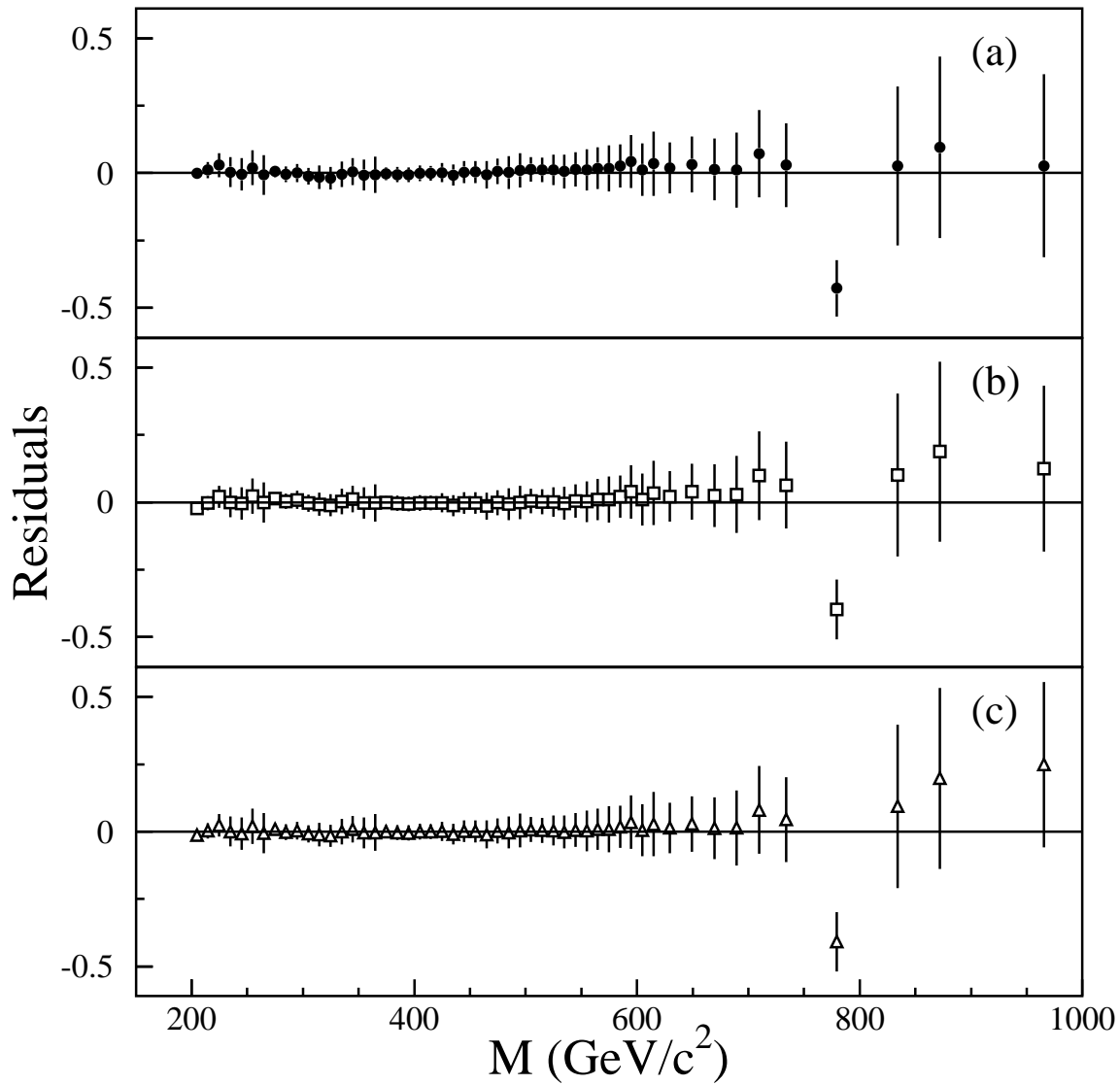,width=6in}}}
\caption{The values of the residuals ([data - fit]/fit) for each of 
 ansatz functions fitted to the data with no signal function. The
 solid circles (a) show the results of the fits using the ansatz of
 equation~\ref{eq_1}, the open squares (b) show the results for
 Equation~\ref{eq_3}, and the open triangles show the results for
 Equation~\ref{eq_4}.}
\label{fig_1a}
\end{figure}

 Three different ansatz functions will be investigated. The first is
\begin{equation}  
F_{i} = \int_{M_{\mbox{\rm min}}}^{M_{\mbox{\rm max}}} A
\left({\jjmass \over 100}\right)^{- \alpha}
\left(  1 - { \jjmass \over 1800 } \right)^{-\beta} d\jjmass .
\label{eq_1}
\end{equation}  
 where A is the normalization parameter and $\alpha$ and $\beta$ are
 fit parameters. Two additional ansatz functions were investigated. The first
 (equation~\ref{eq_3}) was used by the CDF
 collaboration~\cite{cdfxjj2} and the second (equation~\ref{eq_4}) by
 the UA2 collaboration~\cite{UA2wjj,ua2xjj}:
\begin{eqnarray}
F_{i} & = & \int_{M_{\mbox{\rm min}}}^{M_{\mbox{\rm
max}}} A \left({\jjmass \over 100}\right)^{- \alpha}
\left[  1 - { \jjmass \over 1800 } 
+ \gamma \left({\jjmass \over 1800}\right)^{2} \right]^{-\beta} d\jjmass ,
\label{eq_3}\\
\nonumber \\
F_{i} & = & \int_{M_{\mbox{\rm min}}}^{M_{\mbox{\rm
 max}}}
A \left({\jjmass \over 100}\right)^{- \alpha} \exp{\left[-\beta
\left({\jjmass \over 100}\right) -\gamma \left({\jjmass \over 100}\right)^{2}
\right]} d\jjmass .
\label{eq_4}
\end{eqnarray}
 The ansatz functions were all fitted to the 100 simulated spectra
 using a binned maximum likelihood method and the {\sc
 minuit}~\cite{minuit} package. The resulting average \chisq\ for the
 fits is 59 for 52 data bins. The residuals ([data - fit]/fit) for
 each of the fits are depicted in Fig.~\ref{fig_1a}. These residuals
 show that the ansatz functions do represent the mass spectra well with
 no obvious biases.

 The 100 simulated spectra were then fitted with a linear combination
 of the ansatz function and a signal line shape, 
\begin{equation}
F_{i} = \left[ \int_{M_{\mbox{\rm min}}}^{M_{\mbox{\rm max}}} A
\left({\jjmass \over 100}\right)^{- \alpha}
\left(  1 - { \jjmass \over 1800 } \right)^{-\beta} d\jjmass \right]
+ N_{X}S_{i}(M_{\qstar}).
\label{eq_1b}
\end{equation}
 where $N_{\rm X}$ is the number of signal events expected from a
 hypothetical particle for the total luminosity and $S_{i}(\qstar)$ is
 the fraction of signal events in a given mass bin $i$ for a given
 signal mass. For this analysis the signal line shape is given by that
 of an excited quark ($\qstar$)~\cite{excited_quarks} which decays to
 a quark and a gluon ($\qstar \rightarrow qg$). The coupling
 parameters of the excited quark theory were set equal to one ($f =
 f^{\prime} = f_{s} = 1.0$) and the compositeness scale was set equal
 to the mass of the excited quark ($\Lambda^{\ast} = M_{\qstar}$). The
 $\qstar$ were simulated with masses from 200 to 975 GeV/$^2$ at 25
 \gev\ intervals using the {\sc pythia}~\cite{pythia} event
 generator. The resulting particle jets were smeared with the assumed
 jet resolutions. Each \qstar\ sample contains fifty thousand events
 . Examples of \qstar\ mass spectra are shown in Fig~\ref{fig_1}.

\begin{figure}
\vbox{\centerline{\psfig{figure=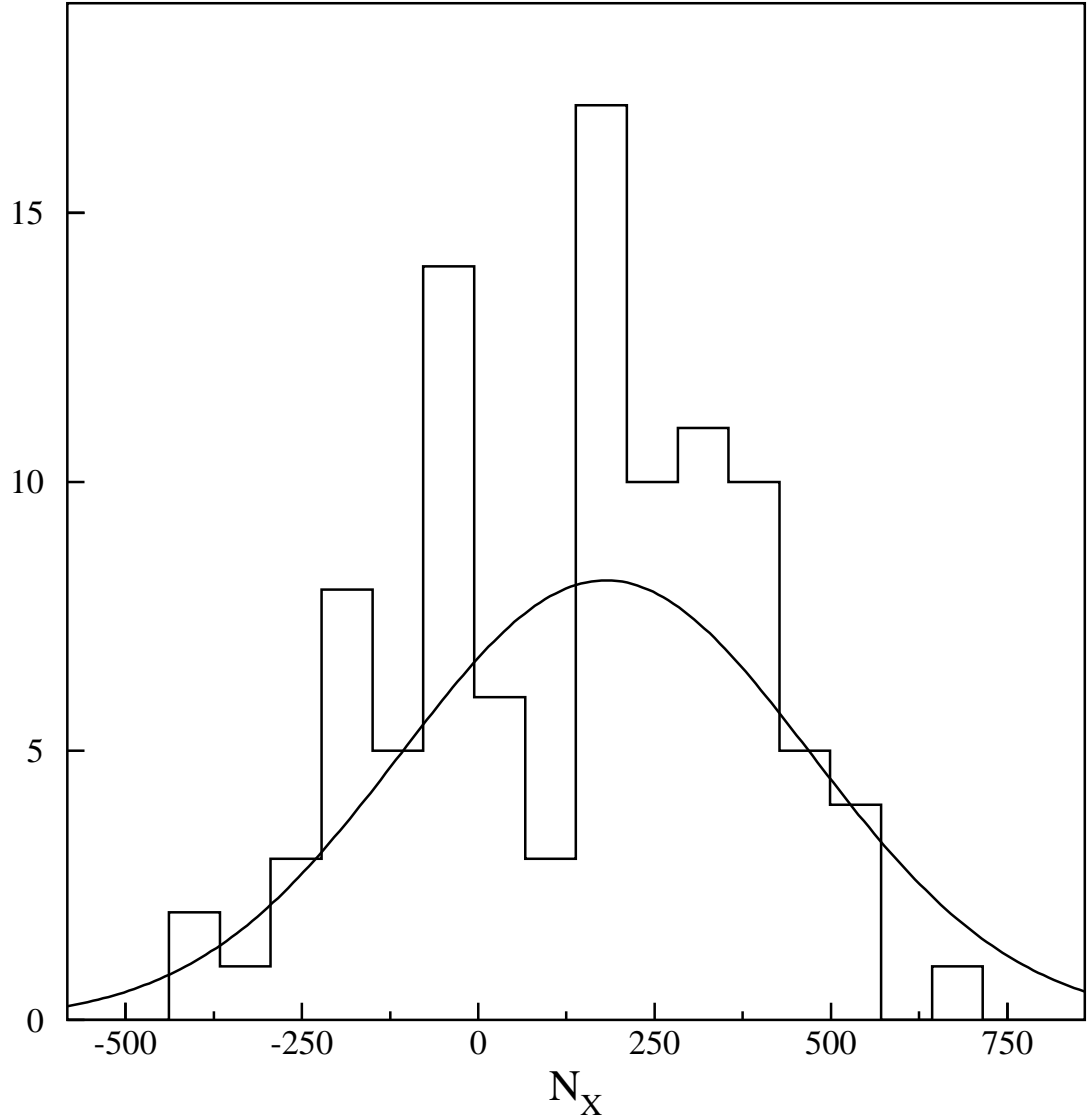,width=6in}}}
\caption{The distribution of the values of $N_{\rm X}$ 
obtained when the 100 simulated mass spectra are fitted with the
ansatz function (Equation~\ref{eq_1}) and a 500 GeV/$^2$ \qstar .}
\label{fig_2}
\end{figure}

 For each of the one hundred background spectra the signal size
 $N_{\rm X}$ and its error $\Delta N_{\rm X}$ was determined for each
 of the \qstar\ masses generated. The average values of $N_{\rm X}$
 ($\overline{ N_{\rm X}}$) and $\Delta N_{\rm X}$ ($\overline{\Delta
 N_{\rm X}}$) were then calculated by fitting the distribution of
 $N_{\rm X}$ values obtained from the 100 spectra to a Gaussian
 distribution (see Fig.~\ref{fig_2}). $\overline{ N_{\rm X}}$ is given
 by the central value of the Gaussian and $\overline{\Delta N_{\rm
 X}}$ is given by the width of the Gaussian.
 
 The 95$\%$ confidence limit (CL) on the excited quark production
 cross section was then determined by assuming that the probability
 density as a function of cross section is given by a Gaussian with a
 center $\overline{ N_{\rm X}}/{\cal{L}}$ and width $\overline{\Delta
 N_{\rm X}}/{\cal{L}}$. The $95\%$ CL on the cross section
 ($\sigma_{\mbox{\rm 95\%}}$) is then given by the value
 of the cross section such that $95\%$ of the physical part ({\rm
 i.e.} the cross section is greater than zero) of the probability
 density function is below this value.

\begin{figure}
\vbox{\centerline{\psfig{figure=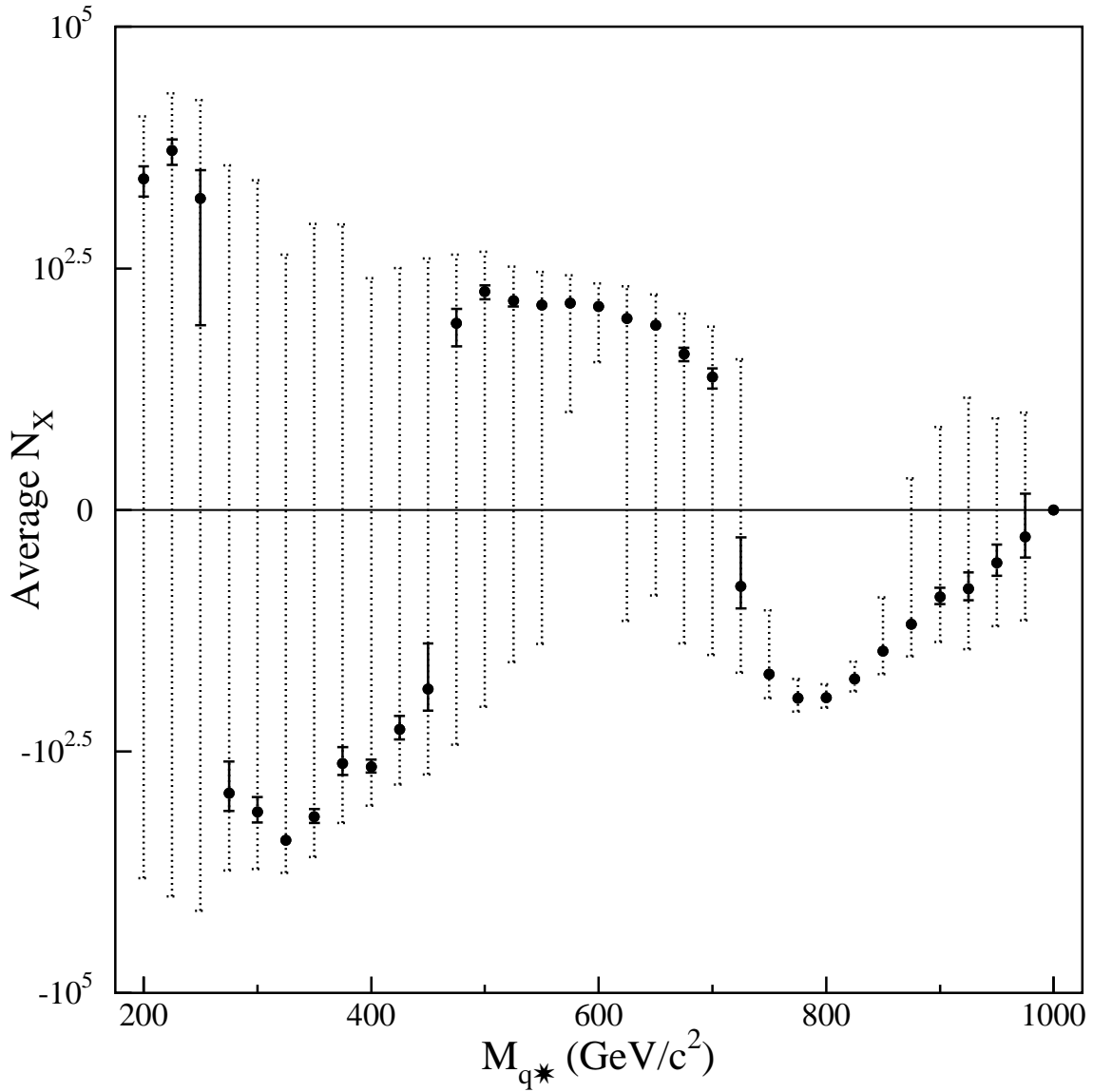,width=6in}}}
\caption{The values of $\overline{N_{\rm X}}$ (solid circles), the 
uncertainty of $\overline{N_{\rm X}}$=$\overline{\Delta N_{\rm
X}}/\sqrt{99}$ (solid error bars) and $\overline{\Delta N_{\rm X}}$
(dotted error bars) obtained from fitting the simulated dijet mass
spectra with Equation~\ref{eq_1}.}
\label{fig_3}
\end{figure}

 Since the spectra being fitted are derived from a NLO QCD
 calculation, the value of $\overline{N_{\rm X}}$ should be consistent
 with zero. Any deviation is the result of an inherent bias in the
 ansatz method. Figure~\ref{fig_3} shows the fitted $\overline{N_{\rm
 X}}$ versus dijet mass. The error on $\overline{N_{\rm X}}$ is taken
 to be $\overline{\Delta N_{\rm X}}/\sqrt{99}$, and it is clear that
 there are systematic biases in the value of $\overline{N_{\rm X}}$ as
 a function of signal (\qstar ) mass.

\begin{figure}
\vbox{\centerline{\psfig{figure=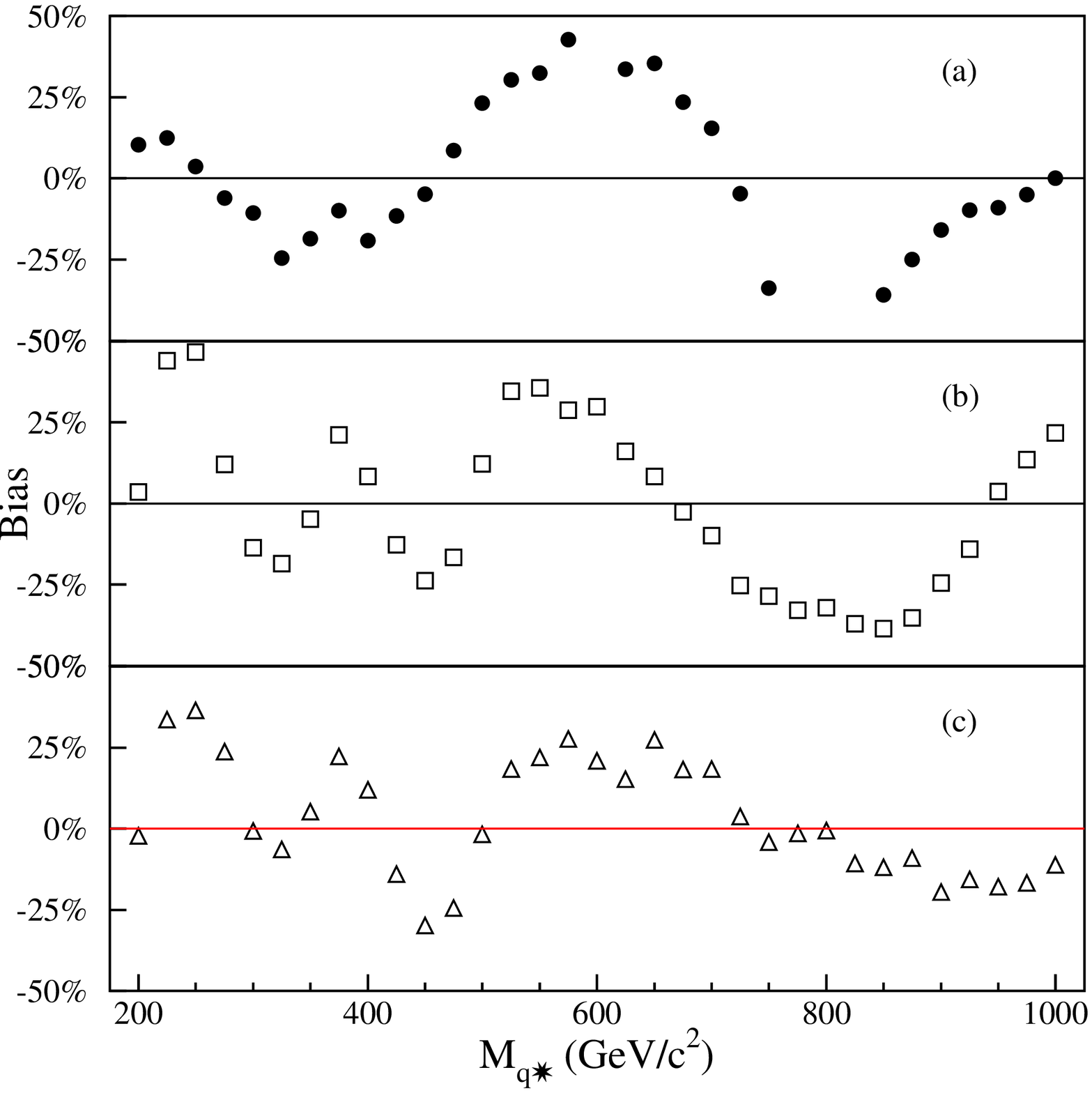,width=6in}}}
\caption{The Bias in the ansatz function method. The solid circles (a) show 
the bias due to ansatz of Equation~\ref{eq_1}, the open squares (b)
show the bias due to the ansatz of Equation~\ref{eq_3} and the open
triangles (c) show the bias due to the ansatz of Equation~\ref{eq_4}.}
\label{fig_4}
\end{figure}

 If there were no bias in the ansatz fitting method the $95\%$ CL on
 the production cross section would be given by $1.96 \times
 \overline{\Delta N_{\rm X}}$. A measure of the bias in the ansatz is
 the the difference between $\sigma_{\mbox{\rm 95\%}}$ and $1.96
 \times \overline{\Delta N_{\rm X}}$:
\begin{equation}
\mbox{Bias} = { {\sigma_{\mbox{\rm 95\%}} - 
(1.96 \times \overline{\Delta N_{\rm  X}})} 
\over {1.96 \times \overline{\Delta N_{\rm X}}} }
\label{eq_10}
\end{equation}
 The resulting biases are given in Table~\ref{table_1} and are
 depicted in Figure \ref{fig_4} (solid circles). The plot clearly
 shows that the resulting $95\%$ CL on the production cross section
 can be miss-measured by up to $\pm 50\%$.

 To see if these biases are caused by the specific choice of ansatz
 function, the two other possibilities were examined
 (equation~\ref{eq_3} and \ref{eq_4}). The bias values for these
 ansatz functions are given in Table~\ref{table_2} and are plotted in
 Fig.~\ref{fig_4}. It is clear that all three of the ansatz functions
 investigated produce a large bias in the resulting $95\%$ CL on the
 cross-section of up to 50$\%$. For a large range of mass values all
 of the ansatz functions produce biases of the same sign and
 approximate value and in some cases none of the limits will include
 the true 95$\%$ CL on the cross section. As limits are being set on
 new particle it is important to note that for a significant number of
 the mass values investigated the ansatz function method
 underestimates the $95\%$ CL.
 
 The bias in the ansatz function method is caused by the changing
 slope of the dijet mass spectrum and the presence of a parameterized
 signal function. This signal function gives the ansatz function a
 flex point at which it can adapt to the changing slope of the data.
 The signal line shape fills a gap between the ansatz and the data
 producing a low $\chi^2$ fit. Therefore the requirement that the
 ansatz function fits the data with a small $\chi^2$ is not sufficient
 to show that the method is unbiased. It is also necessary to show
 that the method does not find a nonexistent signal.

 The systematic uncertainties reported in previous
 searches~\cite{UA2wjj,ua2xjj,cdfxjj,cdfxjj2} are 50 to 300$\%$
 depending on mass of the hypothetical particle. Hence, the limits
 reported by these searches will not be significantly degraded by this
 additional uncertainty. The bias in the method will prevent
 significant reduction of the systematic uncertainties in the
 future. The alternative to using an ansatz function to represent the
 background in these searches is a theoretical prediction (for example
 {\sc jetrad}). Currently the uncertainties in these predictions are
 30--40$\%$~\cite{inc_jet_theory_uncertainties} due to choice of
 renormalization scale and pdf. The effect of this uncertainty will
 have to be included in any future searches (this uncertainty can be
 reduced by improvements in the accuracy of pdf's and theoretical
 calculations).

 In conclusion it has been shown that the ansatz method of searching
 for new particles has an inherent bias in the method and this bias
 has to be accounted for when placing limits on the production
 cross-sections of new particle production.

 I thank my colleagues on the D\O\ experiment for their helpful
 comments, suggestions and discussions. {\em This work was supported
 in part by the Department of Energy under grant DE-FG02-91ER40684 at
 Northwestern University.}

\begin{table*}[htbp]
\begin{center}
\caption{The $95\%$ Confidence Limits obtained as a function of the mass 
of the \qstar . Presented are the limits calculated from the fit using
$\overline{N_{\rm  X}}$ and $\overline{\Delta N_{\rm
 X}}$, the limit calculated using $1.96 \times
\overline{\Delta N_{\rm  X}}$ and the bias.}
\vspace{2mm}
\label{table_1}
\begin{tabular}{dd@{$\times 10$}ld@{$\times 10$}lddd@{$\times 10$}ld@{$\times 10$}ld}
\multicolumn{1}{c}{$\qstar$} & \multicolumn{2}{c}{Width} & 
\multicolumn{2}{c}{Fit}   & \multicolumn{1}{c}{Bias}  &
\multicolumn{1}{c}{$\qstar$} & \multicolumn{2}{c}{Width} & 
\multicolumn{2}{c}{Fit}   & \multicolumn{1}{c}{Bias}  \\
\multicolumn{1}{c}{Mass} & \multicolumn{2}{c}{Limit}& 
\multicolumn{2}{c}{Limit} & \multicolumn{1}{c}{}  &
\multicolumn{1}{c}{Mass} & \multicolumn{2}{c}{Limit}& 
\multicolumn{2}{c}{Limit} & \multicolumn{1}{c}{}  \\
\multicolumn{1}{c}{} & 
\multicolumn{2}{c}{$1.96 \times \overline{\Delta N_{\rm X}}$}& 
\multicolumn{2}{c}{$\sigma_{\mbox{\rm 95\%}}$} & \multicolumn{1}{c}{}  &
\multicolumn{1}{c}{} & 
\multicolumn{2}{c}{$1.96 \times \overline{\Delta N_{\rm X}}$}& 
\multicolumn{2}{c}{$\sigma_{\mbox{\rm 95\%}}$} & \multicolumn{1}{c}{}  \\
\multicolumn{1}{c}{(\gev )} & \multicolumn{2}{c}{($pb$)} &
\multicolumn{2}{c}{($pb$)}  & \multicolumn{1}{c}{($\%$)} &
\multicolumn{1}{c}{(\gev )} & \multicolumn{2}{c}{($pb$)} &
\multicolumn{2}{c}{($pb$)}  & \multicolumn{1}{c}{($\%$)} \\
\hline\hline 
 200. &    1.80 &$^2$&  1.99 &$^2$&   10.   & 
 225. &    2.98 &$^2$&  3.35 &$^2$&   12.   \\
 250. &    3.11 &$^2$&  3.22 &$^2$&    4.   & 
 275. &    8.88 &$^1$&  8.35 &$^1$&   -6.   \\
 300. &    7.64 &$^1$&  6.82 &$^1$&  -11.   & 
 325. &    6.00 &$^1$&  4.53 &$^1$&  -24.   \\
 350. &    4.73 &$^1$&  3.85 &$^1$&  -18.   & 
 375. &    2.59 &$^1$&  2.33 &$^1$&  -10.   \\
 400. &    1.38 &$^1$&  1.11 &$^1$&  -19.   & 
 425. &    9.92 &$^0$&  8.78 &$^0$&  -12.   \\
 450. &    9.27 &$^0$&  8.82 &$^0$&   -5.   & 
 475. &    6.92 &$^0$&  7.51 &$^0$&    8.   \\
 500. &    5.69 &$^0$&  7.01 &$^0$&   23.   & 
 525. &    3.60 &$^0$&  4.70 &$^0$&   30.   \\
 550. &    3.07 &$^0$&  4.07 &$^0$&   32.   & 
 575. &    2.51 &$^0$&  3.59 &$^0$&   43.   \\
 600. &    1.84 &$^0$&  2.86 &$^0$&   56.   & 
 625. &    2.16 &$^0$&  2.88 &$^0$&   34.   \\
 650. &    1.74 &$^0$&  2.36 &$^0$&   35.   & 
 675. &    1.28 &$^0$&  1.59 &$^0$&   24.   \\
 700. &    1.09 &$^0$&  1.26 &$^0$&   15.   & 
 725. &    0.83 &$^0$&  0.79 &$^0$&   -5.   \\
 750. &    0.76 &$^0$&  0.51 &$^0$&  -34.   & 
 775. &    0.65 &$^0$&  0.29 &$^0$&  -55.   \\
 800. &    0.47 &$^0$&  0.17 &$^0$&  -64.   & 
 825. &    0.38 &$^0$&  0.16 &$^0$&  -57.   \\
 850. &    0.41 &$^0$&  0.26 &$^0$&  -36.   & 
 875. &    0.34 &$^0$&  0.26 &$^0$&  -25.   \\
 900. &    0.30 &$^0$&  0.25 &$^0$&  -16.   & 
 925. &    0.41 &$^0$&  0.37 &$^0$&  -10.   \\
 950. &    0.24 &$^0$&  0.22 &$^0$&   -9.   & 
 975. &    0.24 &$^0$&  0.23 &$^0$&   -5.   \\
\end{tabular}
\end{center}
\end{table*}

\begin{table*}[htbp]
\begin{center}
\caption{The Biases for all three ansatz functions investigated.
(Equation~\ref{eq_1}, Equation~\ref{eq_3} and Equation~\ref{eq_4} )}
\vspace{2mm}
\label{table_2}
\begin{tabular}{dddddddd}
\multicolumn{1}{c}{$\qstar$} & \multicolumn{3}{c}{Bias}  &
\multicolumn{1}{c}{$\qstar$} & \multicolumn{3}{c}{Bias} \\
\cline{2-4}\cline{6-8}
\multicolumn{1}{c}{Mass} & \multicolumn{1}{c}{Equation~\ref{eq_1}} &
\multicolumn{1}{c}{Equation~\ref{eq_3}} &\multicolumn{1}{c}{Equation~\ref{eq_4}} &
\multicolumn{1}{c}{Mass} & \multicolumn{1}{c}{Equation~\ref{eq_1}} &
\multicolumn{1}{c}{Equation~\ref{eq_3}} &\multicolumn{1}{c}{Equation~\ref{eq_4}} \\
\multicolumn{1}{c}{(\gev )} & \multicolumn{1}{c}{($\%$)} &
\multicolumn{1}{c}{($\%$)}   & \multicolumn{1}{c}{($\%$)} &
\multicolumn{1}{c}{(\gev )} & \multicolumn{1}{c}{($\%$)} &
\multicolumn{1}{c}{($\%$)}   & \multicolumn{1}{c}{($\%$)} \\
\hline \hline
 200. &      10.   &       4.   &      -2.   & 
 225. &      12.   &      44.   &      34.   \\
 250. &       4.   &      47.   &      36.   & 
 275. &      -6.   &      12.   &      24.   \\
 300. &     -11.   &     -14.   &      -1.   & 
 325. &     -24.   &     -19.   &      -6.   \\
 350. &     -18.   &      -5.   &       5.   & 
 375. &     -10.   &      21.   &      22.   \\
 400. &     -19.   &       8.   &      12.   & 
 425. &     -12.   &     -13.   &     -14.   \\
 450. &      -5.   &     -24.   &     -30.   & 
 475. &       8.   &     -17.   &     -24.   \\
 500. &      23.   &      12.   &      -2.   & 
 525. &      30.   &      35.   &      18.   \\
 550. &      32.   &      36.   &      22.   & 
 575. &      43.   &      29.   &      28.   \\
 600. &      56.   &      30.   &      21.   & 
 625. &      34.   &      16.   &      15.   \\
 650. &      35.   &       8.   &      27.   & 
 675. &      24.   &      -2.   &      18.   \\
 700. &      15.   &     -10.   &      18.   & 
 725. &      -5.   &     -25.   &       4.   \\
 750. &     -34.   &     -28.   &      -4.   & 
 775. &     -55.   &     -33.   &      -2.   \\
 800. &     -64.   &     -32.   &       0.   & 
 825. &     -57.   &     -37.   &     -11.   \\
 850. &     -36.   &     -38.   &     -12.   & 
 875. &     -25.   &     -35.   &      -9.   \\
 900. &     -16.   &     -24.   &     -20.   & 
 925. &     -10.   &     -14.   &     -16.   \\
 950. &      -9.   &       4.   &     -18.   & 
 975. &      -5.   &      13.   &     -17.   \\
\end{tabular}
\end{center}
\end{table*}

\begin{thebibliography}{99}

\bibitem{UA2wjj}
J.~Alitti {\rm et al.} (UA2 Collaboration), Zeit. Phys. C {\bf 49}, 17 1991.

\bibitem{ua2xjj}
J.~Alitti {\rm et al.} (UA2 Collaboration), Nucl. Phys. {\bf B400}, 3
(1993).

\bibitem{cdfxjj}
F. Abe {\rm et al.} (CDF Collaboration), Phys. Rev. Lett. {\bf 74}, 3538 (1995), hep-ex/9501001. 

\bibitem{cdfxjj2}
F. Abe {\rm et al.} (CDF Collaboration), Phys. Rev. D {\bf 55}, (1997), hep-ex/9702004.

\bibitem{jetrad} 
W.T.~Giele, E.W.N. Glover and D.A. Kosower, Nucl. Phys. {\bf B403},
633 (1993).

\bibitem{cteq} 
H.L. Lai {\rm et al.}, Phys. Rev. D {\bf 55}, 1280 (1997),
hep-ph/9606399.

\bibitem{rsep} 
B. Abbott {\em et al.} (for the D\O\ Collaboration), 
Fermilab-Pub-97/242-E.

\bibitem{jet_resolutions}
 M. Bhattacharjee, {\em et al.}, D\O\ Internal Note 2887, {\it Jet
 Energy Resolution}, ( May 22, 1996)

\bibitem{dijet_mass} B.~Abbott et al. (D\O\ Collaboration),
   hep-ex/9807014, submitted to Phys. Rev. Lett.


\bibitem{minuit} 
 F. James, CERN Program Library Entry D506 (unpublished),\\
 {\tt http://wwwinfo.cern.ch/asdoc/minuit\_html3/minmain.html.}\\
 {\sc minuit} version 96.a was used.

\bibitem{excited_quarks} U.~Baur, M.~Spira and P.M. Zerwas, Phys. Rev. D {\bf 42}, 815 (1990).

\bibitem{pythia} 
 T. Sj\"ostrand, Computer Physics Commun. {\bf 82}, 74 (1994).  {\sc
 pythia} version 5.7 was used.

\bibitem{inc_jet_theory_uncertainties}
B.~Abbott, {\em et al.},  hep-ph/9801285, 
Eur.\ Phys.\ J. C {\bf 5} 687 (1998).


\end{thebibliography}
\end{document}